\documentstyle[12pt]{article}
\def\beq{\begin{equation}}
\def\eeq{\end{equation}}
\def\bea{\begin{eqnarray}}
\def\nn{\nonumber \\ }
\def\eea{\end{eqnarray}}

\def\ds{\displaystyle}

\def\req#1{(\ref{#1})}

\begin{document}

\thispagestyle{empty}
\hfill{CPTH-A334.1094}\\
\phantom{bla}
\hfill{   }\\
\phantom{bla}
\hfill{October 1994}
\vskip 0.5cm
\begin{center}
{\Large Renormalization group and logarithmic corrections \\
to scaling relations in conformal sector of 4D gravity}\\
\vspace{1cm}
{\bf I. Antoniadis}\\
\vspace{0.5cm}
Centre de Physique Th{\'e}orique
\footnote{Laboratoire Propre du CNRS UPR A.0014}\\
Ecole Polytechnique, 91128 Palaiseau Cedex, France\\
\vspace{0.5cm}
{\bf and}\\
\vspace{0.5cm}
{\bf Sergei D. Odintsov,} \\
\vspace{0.5cm}
Tomsk Pedagogical Institute, 634041 Tomsk, Russia\\
and   Dept. ECM, Faculty of Physics \\
Diagonal 647, Universidad de Barcelona, 08028 Barcelona, Spain \\
\end{center}
 \vspace{2cm}

\abstract{We study the effective theory of the conformal factor
near its infrared stable fixed point. The renormalization group
equations for the effective coupling constants are found and
their solutions near the critical point are obtained, providing the
logarithmic corrections to the scaling relations. Some
cosmological applications of the running of coupling constants are
briefly discussed.}

\newpage
It is quite an old idea to work with some effective theory for
 quantum gravity (QG) in the absence of a consistent theory. The
traditional candidate for such an effective theory is usually Einstein
 gravity which is known to be non-renormalizable \cite{1}.
Despite this fact, already in this theory the loop corrections can be
calculated and may become quite relevant for a variety of phenomena
\cite{2,3} (for a general introduction to perturbative QG, see for example
\cite{4}).
   Recently, an interesting effective model aiming to describe
 QG in the far infrared, has been introduced \cite{5}. This model is based on
the
 effective theory for the conformal factor, and some of its properties have
been
 further studied in refs.\cite{6}-\cite{9}. It was found that it possesses a
non-trivial infrared (IR) stable fixed point which could become physically
relevant at cosmological distance scales. Investigation of this critical
point led
 to the derivation of exact scaling relations and, in particular, of the
 anomalous dimension of the conformal factor in analogy with 2D
non-critical string theory. Furthermore, this model provides a dynamical
framework for a solution of
 the cosmological constant problem \cite{5}.

The purpose of this letter is to study the effective theory of
the conformal factor near its fixed point. We show that making the
one-loop renormalization in a way which preserves global conformal
invariance of counterterms, one can rigorously construct
renormalization group (RG) equations for coupling constants. The solution of
these equations near the critical point gives the logarithmic corrections to
scaling relations.

Consider the flat space theory described by the Lagrangian
\beq
{\cal L}= b_1 \left( \Box
\sigma \right)^2 +  b_2 \left( \partial_\mu \sigma
\right)^2 \Box \sigma +  b_3  \left[ \left( \partial_\mu \sigma
\right)^2 \right]^2
 + b_4 (\sigma ) \left( \partial_\mu \sigma \right)^2 +  b_5 (\sigma )
\eeq
where $\sigma$ has classically canonical dimension zero, $b_1,b_2,b_3$ are
dimensionless coupling constants and $b_4,b_5$ are for the moment arbitrary
dimensionful field-dependent generalized couplings. The one-loop divergences
of this theory were computed in ref.\cite{9} and for some choices of
$b_4,b_5$ it is multiplicatively renormalizable in the
usual sense. These choices correspond to the effective theory
of the conformal factor \cite{5} which we review below.

We start for simplicity with $N$ conformally invariant scalars in curved
spacetime. The conformal anomaly in this case is  \cite{10}
\beq
T_{\mu}^{\mu}=b\left( F+{2 \over 3}\Box R \right)
+b'G + b''\Box R,
\eeq
where
\[ b={ N \over 120 (4 \pi^2) }, \ \ b'=-{ N \over 360 (4 \pi^2) },  \]
and $b''$ may be changed by the variation of a local $R^2$ counterterm in the
gravitational part of the action. In eq.(2) $R$ is the scalar curvature, $F$
is the square of Weyl tensor and $G$ is the Gauss-Bonnet combination.
Choosing the conformal parametrization $g_{\mu\nu}(x)=e^{2\sigma}\bar
g_{\mu\nu}$
and integrating over $\sigma$, we get an anomaly induced action
\cite{11,5}. Adding to this action the classical Einstein action
in conformal parametrization we
get the effective theory for the conformal factor.
In notations of ref.\cite{5} and in flat background metric
${\bar g_{\mu\nu}}=\eta_{\mu\nu}$, the resulting Lagrangian is
\beq
\ds {\cal L} = \ds-{ Q^2 \over (4\pi)^2 } (\Box\sigma)^2
-\zeta\left[ 2\alpha(\partial_{\mu}\sigma)^2 \Box\sigma
+\alpha^2 (\partial_{\mu}\sigma)^4 \right] \nn
+ \gamma e^{2 \alpha \sigma} (\partial_{\mu}\sigma)^2
-{\lambda \over \alpha^2} e^{4 \alpha \sigma},
\label{Lcf}
\eeq
where
$\zeta=b+2b'+3b''$, $\ds{ Q^2 \over (4\pi)^2 }=\zeta-2b'$,
$\ds\gamma ={3 \over \kappa}$ and $\lambda ={\Lambda \over \kappa}$ with
$\kappa =8\pi G$. $\zeta$ is the coupling of the local $R^2$ term, $\gamma$
and $\lambda$ are the Newtonian and the cosmological couplings,
respectively, while $Q^2$ at
$\zeta=0$ was interpreted as a four-dimensional central charge for which
a possible four-dimensional C-theorem could be applied \cite{6,12}. In
eq.(3) we also allowed for a non-trivial anomalous scaling dimension
$\alpha$ for the conformal factor $e^{\alpha\sigma}$; its classical value is
$\alpha=1$.

As it can be easily seen, the theory (3) belongs to the type
(1) and it is multiplicatively renormalizable by simple power counting
arguments. It has been studied in the IR stable fixed point
$\zeta=0$ where it was argued that it describes an infrared phase of
QG \cite{5}. Here, we will discuss the behavior of the effective couplings
away from the fixed point in a spirit different from the discussion of
ref.\cite{9}, using a renormalization procedure based on physical
requirements. In fact, the Lagrangian (3) is the most general one in flat
space, containing up to 4 space-time derivatives and being invariant under
global conformal transformations, with the conformal factor
$e^{\alpha\sigma}$ having canonical dimension 1 \cite{5}. Global conformal
symmetry is actually the remnant of full general covariance of the effective
theory, once the background metric is fixed to the flat metric.

An inspection of the structure of divergences based on power counting
arguments shows that the four-derivative sector gets renormalized by
diagrams involving only $\zeta$-vertices. Moreover, there are in general
only two independent renormalizations which can be chosen to be those of
$\zeta$ and
$\alpha$. There is no independent wave function renormalization because
the coefficient $b'$ multiplies a non-local term in the action when
expressed in terms of the full metric. On the other hand, the
two-derivative coupling $\gamma$ and the cosmological coupling $\lambda$
are renormalized by diagrams involving in addition to $\zeta$-vertices,
exactly one $\gamma$-vertex, and two $\gamma$-vertices or one
$\lambda$-vertex, respectively. A simple counting of combinatoric factors
of higher-point functions then shows that all such divergences are of the
form $e^{2\alpha\sigma} (\partial_\mu\sigma)^2$ and $e^{4\alpha\sigma}$. It
follows that theory (3) is renormalizable by redefining the couplings
$\zeta$, $\gamma$, $\lambda$ and $\alpha$.

The one-loop counterterms for the theory (1) have been calculated in
ref.\cite{9}. Specifying the results to the model (3), we get the following
expression for the corresponding one-loop divergences (in dimensional
regularization):
\beq
\begin{array}{ll}
\ds\Gamma_{\rm div}=-{2 \over \varepsilon} \int d^4x \sqrt{-g}
&\ds{ (4\pi)^4 \over Q^4 } \ds\left\{
5 \alpha^2 \zeta^2 \left[ \Box\sigma
+\alpha (\partial_{\mu}\sigma)^2 \right]^2
\right. \\
&\ds\left. - \gamma \alpha^2
\left( 3 \zeta + { 2Q^2 \over (4\pi)^2 } \right)
(\partial_\mu\sigma)^2 e^{2 \alpha\sigma} +
 \left( {8 \lambda Q^2 \over (4\pi)^2} -{\gamma^2 \over 2} \right)
e^{4 \alpha\sigma} \right\}
\end{array}\label{Gdiv}
\eeq
It follows from expressions \req{Gdiv} and \req{Lcf}, that one-loop
renormalization indeed preserves global conformal invariance in the
structure of counterterms. Furthermore, in the four-derivative sector, only
the coupling constant $\zeta$ should be renormalized at the one-loop
level, while $\alpha$ remains a free parameter.
The corresponding $\beta$-functions are:
\beq
\begin{array}{lll}
\beta_{\zeta}&=&\ds {80 \pi^2 \alpha^2 \zeta^2 \over Q^4} \\
\beta_{\gamma}&=&\ds \left[ {2\alpha^2 \over Q^2}
+3\zeta\alpha^2 {(4\pi)^2 \over Q^4} \right] \gamma \\
\beta_{\lambda}&=&\ds {8\lambda\alpha^2 \over Q^2}
-{8 \pi^2 \alpha^2 \gamma^2 \over Q^4} \ .
\end{array}
\eeq

Now let us write the RG equations for the effective action. Their general
form is:
\beq
\left\{ {\partial \over \partial t}
- (\beta_p+pd_p){\partial \over \partial p}
-(\gamma_{\sigma}+d_{\sigma}) \int d^n x
\sigma(x){ \delta \over \delta \sigma(x)} \right\}
\Gamma[ e^t k, \sigma, p, \mu ]=0,
\eeq
where $\Gamma$ is the renormalized  effective action,
$p$ stands for all the renormalized couplings $\{ \zeta, \gamma, \lambda
\}$, $d_p$ and $d_{\sigma}$ are the classical scaling dimensions for $p$ and
$\sigma$, and $\gamma_{\sigma}$ is the anomalous dimension of $\sigma$. As
we have seen above, $d_{\sigma}=\gamma_{\sigma}=0$. Finally $k$ denotes the
external momenta, and $t \to \infty$ ($ -\infty$) corresponds to the
ultraviolet (infrared) limit.

The RG equations for the effective couplings (which determine the
asymptotic behavior of the effective action $\Gamma$) are:
\beq
{d p(t) \over dt}=\beta_p(t)+d_p p(t)
\eeq
Using eq.(5) and the classical scaling dimensions of the various couplings
$p$ \cite{5}:
$d_{\zeta}=0$, $d_{\gamma}=2-2\alpha$, $d_{\lambda}=4-4\alpha$, the
one-loop RG equations for the effective couplings become:
\bea
\ds{d \zeta(t) \over dt}&=&\ds {80 \pi^2 \alpha^2 \zeta^2(t) \over Q^4(t)}
,\\
\ds{d\gamma(t) \over dt}&=&\ds\left[  {2 \alpha^2 \over Q^2(t)}
+3 \zeta(t) \alpha^2 {(4\pi)^2 \over Q^4(t)} +2 -2\alpha \right]
\gamma(t), \\
\ds{d\lambda(t) \over dt}&=&\ds\left[
{8\lambda(t)\alpha^2 \over Q^2(t)}
-{8 \pi^2 \alpha^2 \gamma^2(t)\over Q^4(t)} +(4 -4\alpha) \lambda(t)
\right].
\label{RGeqeffc}
\eea
We remind that $Q^2(t)$ is not an independent function:
$\ds {Q^2(t) \over (4 \pi)^2}=\zeta(t)-2b'$. It is also useful to define
the dimensionless ratio $h\equiv\lambda/\gamma^2$ which satisfies the
equation:
\beq
{d h(t) \over dt}= {4 \alpha^2 \over Q^2(t)}h(t) -
{8 \pi^2 \alpha^2 \over Q^4(t)}-
6{\alpha^2 (4\pi)^2 \over Q^4(t)} \zeta(t) h(t) \ .
\eeq

The system of eqs.(8)-(11) was studied in the IR stable fixed point
$\zeta=0$ in ref.\cite{5}. The anomalous scaling dimension of the
conformal factor is determined by the vanishing of (9), while the fixed
point of (11) determines the relation between the cosmological and
Newtonian couplings. At one-loop level, the IR fixed point solutions are:
\beq
\zeta=0, \ \
\alpha =\alpha_0 \equiv{ 1 - \left( 1-{4 \over Q_0^2} \right)^{1/2} \over
2/Q_0^2},\ \
h = { 2 \pi^2 \over Q_0^2 },\ \ \gamma={\rm arbitrary},
\eeq
where $Q_0^2\equiv Q^2(\zeta=0)$.

Now, the solution of (8) can be found only in non-explicit form:
\beq
\zeta^2-4b'\zeta \log \zeta-4b^{\prime 2}=
{5 \alpha^2 \zeta  \over (4 \pi)^2}t + c\zeta
\eeq
where c is an integration constant. The form
of (13) is not very useful for obtaining analytic expressions for the
solutions of RG equations. However, here we are interested in the
asymptotic behavior of the solutions near the infrared fixed point (12). In
this limit, eq.(13) gives:
\bea
\zeta(t)&\simeq&-{ 4b^{\prime 2}(4 \pi)^2 \over 5\alpha^2 t }\ \ \to 0 \nn
t&\to&-\infty
\eea
This is an asymptotically free solution in the infrared; its presence was
also discussed in ref.\cite{9}, using a different form of renormalization
of coupling constants. Using eq.(14), the solution of (9) and (11) in the
infrared limit ($t \to-\infty$) is:
\bea
\gamma(t)&\simeq&\ds (-t)^{-1/5}
e^{ t\left( 2 -2\alpha +{2\alpha^2 \over Q_0^2} \right) }, \nn
h(t)&\simeq&\ds {2\pi^2\over Q_0^2} + e^{ {4\alpha^2 \over Q_0^2}t}\ .
\eea

Equations (14) and (15) give the leading logarithmic corrections to coupling
constants in the infrared limit. Using the fixed point value
$\alpha=\alpha_0$ from eq.(12), one finds that the Newtonian coupling
$\gamma$ decreases slowly at large distances as:
\beq \gamma(t)\simeq (-t)^{-1/5}\ . \eeq
This result is specific to one-loop, since $\alpha$ remains
unrenormalized at this level. Inclusion of higher order radiative
corrections is expected in general to destabilize this behavior.
In fact, for $\alpha$ away but close to its fixed point value (12),
one gets an exponential growth or decay of $\gamma$, depending on whether
$\alpha$ approaches $\alpha_0$ from above or from below, respectively.
Therefore, a two-loop computation is needed to clarify this point. On the
other hand, the cosmological coupling behaves in all cases as the square of
the Newtonian coupling. In fact, the ratio $h(t)$ approaches its asymptotic
value exponentially fast.

One may consider a different physical interpretation of the above results,
if one chooses the Planck scale as a unit of mass at all energies.
In this context, the Newtonian coupling should be normalized to one,
implying the vanishing of its derivative (9). In this way, we obtain an
effective anomalous scaling dimension for the conformal factor:
\beq
\alpha(t) \simeq \alpha_0 + 4\pi^2\zeta(t) \left(
-1+ (1-4/Q_0^2)^{1/2} +2 (1-4/Q_0^2)^{-1/2} \right) .
\eeq

The infrared running of the various couplings and in particular of Newton's
constant may have interesting physical applications. Consider for instance
the Newtonian potential which has the form:
\beq V(r)=-{G m_1 m_2 \over r}, \eeq
where $ G=3/ 8\pi\gamma$. The Wilsonian gravitational potential, obtained
from (18) by replacing $G$ with
the running Newton coupling, may have important consequences in
cosmology, in particular to the dark matter problem \cite{13}.
Naturally, one has to use the RG parameter,
$\ds t={1 \over 2}\log{r_0^2 \over r^2}$,
in analogy with the case of electrostatic potential in
quantum electrodynamics
$\ds\left({\mu_0^2 \over \mu^2}\sim {r^2 \over r_0^2} \right)$.
As a result, we get
\beq
V(r)\simeq -{G_0 m_1 m_2 \over r}
\left( 1-{ \alpha^2 \over 10 Q_0^2} \log{r_0^2 \over r^2}
\right) ,
\eeq
where $G_0$ is the (initial) value of the Newton's constant at the distance
$r_0$. Eq.(19) describes the leading-log corrections to the classical
gravitational potential in the context of the effective theory of QG
under investigation. Note that in Einstein action, the
leading-log corrections are absent due to the trivial fact that the only
coupling in this case is the dimensionful Newton's constant
and loop corrections are fall off with powers of $r^2$.

It was suggested in ref.\cite{13} that a possible logarithmic RG evolution
of Newton's constant could lead to a scale dependence of density
fluctuations with interesting effects to the dark
matter problem. For this purpose, one needs the Newtonian coupling to be a
slowly growing function of distance, so that the density parameter
$\Gamma_0$,
\beq \Gamma_0={8 \pi \over 3} {G \rho_m \over H_0^2 }, \eeq
has a similar behavior. In eq.(20) $H_0$ is the current value of the
Hubble parameter, and $\rho_m$ is the baryon matter density at the present
epoch. This analysis was performed in the context of a higher-derivative
$R^2$-gravity theory. In our case, eq.(16) implies that $G\sim G_0
(-t)^{1/5}$ which suggests that the effective theory of conformal factor
might be promising for such a direction.

Finally, let us present briefly a few remarks on the critical behavior of
this system at non-zero temperature.
Using the well-known analogy between finite-temperature and
infrared properties of the corresponding three-dimensional theory, one may
apply the technique of
$\epsilon$-expansion. For this purpose, it is enough to consider only the
four-derivative sector of (3) (associated to dimensionless couplings in
four dimensions), since all massive terms are irrelevant for this
discussion. In $D=(4-\epsilon)$ dimensions, one has to replace the
dimensionless parameters $Q^2$ and $\zeta$ by $Q^2 \mu^{-\epsilon}$ and
$\zeta \mu^{-\epsilon}$ (the dimension of $\sigma$ is not changing). The RG
equation (8) for $\zeta$ is then modified by the addition of a new term
$\epsilon \zeta(t)$ in the right hand side (at the end we put
as usually $\epsilon=1$). The fixed point solutions are defined by
the equality of the new $\beta$-function for $\zeta$ to zero. As a result,
we obtain three fixed points: $\zeta_1=0$, and
$\zeta_{2,3}= 2 b' -5{\alpha}^2 (2 \epsilon (4 \pi)^2)^{-1} \pm
[( b'-5 \alpha^2 (4\epsilon (4 \pi)^2)^{-1} )^{2} -4 b'^2]^{1/2}$.
Stability is imposed by the condition that the derivative of $\zeta$
$\beta$-function should be positive. It turns out that the fixed point
$\zeta_1=0$ is IR stable, which is an indication that at this point the
system undergoes a second order phase transition.

In summary, we discussed the effective theory of the conformal factor
and found the solutions of RG equations near the critical point. As a
result, we obtained the logarithmic corrections to scaling relations in the
asymptotically free IR regime. We also discussed possible interesting
cosmological applications of the running of Newtonian and cosmological
couplings.

A weak point of theory (3) is the approximation of neglecting the
spin-2 graviton modes \cite{5}. It was argued \cite{6} that their
contribution might only lead to finite renormalization of the conformal
anomaly coefficients at the fixed point, and it could be taken into account
afterwards. In fact, the results of our RG consideration indicate that the
effective theory of the conformal factor could be the approximate
infrared limit of a more complete system of RG equations in the
context of some consistent theory of QG where of course all degrees of
freedom are taken into account. It would be interesting to develop such a
point of view further.

This work has been supported in part by MEC(Spain)
                                , in part by ISF project
RI-1000 (Russia), and in part by the EEC contracts SC1-CT92-0792 and
CHRX-CT93-0340.

\end{document}